\documentclass[12pt]{revtex4-2}
\usepackage{amsmath,amsfonts}
\usepackage{amssymb,latexsym}
\usepackage{color}
\usepackage{graphicx}
\usepackage{here}
\usepackage[hang,small,bf]{caption}
\usepackage[subrefformat=parens]{subcaption}

\newcommand{\ket}[1]{|\kern.3ex#1\kern.3ex\rangle}
\newcommand{\bra}[1]{\langle\kern.3ex #1 \kern.3ex|}
\newcommand{\scalar}[2]{\langle\kern.3ex #1 \kern.3ex|\kern.3ex#2\kern.3ex\rangle}

\newcommand{\um}{\mathrm{m}}

\begin{document}

\title{Possible Enhancements of Collective Flow Anisotropy induced by Uncertainty Relation for Fluid Element}%

\author{Gyell Gonçalves de Matos}
\email{gyellgoncalves@gmail.com}
\affiliation{Instituto de F\'{i}sica, Universidade Federal do Rio de Janeiro}
\author{Takeshi Kodama}
\email{kodama.takeshi@gmail.com}
\affiliation{Instituto de F\'{i}sica, Universidade Federal do Rio de Janeiro}
\affiliation{Instituto de F\'{i}sica, Universidade Federal Fluminense}
\author{Tomoi Koide}
\email{tomoikoide@gmail.com}
\affiliation{Instituto de F\'{i}sica, Universidade Federal do Rio de Janeiro}
\date{\today}%
\begin{abstract}
In the stochastic formulation of viscous hydrodynamics, the velocity of a fluid element fluctuates 
satisfying a similar relation to the quantum-mechanical uncertainty relation.
Using a non-relativistic toy model, 
we show that the presence of such a velocity fluctuation increases the local anisotropy of the momentum distributions of produced hadrons,
and thus the collective flow parameters such as $v_2$ is emphasized.
\end{abstract}
\maketitle
%\tableofcontents

\section{Introduction}

After establishing the pictures of the almost ideal fluid and the strongly coupled quarks and gluons in the deconfinement phase (QGP), 
the hydrodynamic description is considered to be an indispensable tool for studying dynamics of the QCD matter produced in the relativistic heavy-ion collisions.
What we observe in experiment is however not such a matter as a fluid (continuum medium) but the momentum distribution of the final observed hadrons.
Thus the fluid described by relativistic hydrodynamics should be mapped into these hadrons.
This mapping is known as the (kinetic) freeze-out process.
It should be emphasized that we cannot determine such observed hadrons only from the behaviors of the fluid where microscopic degrees of freedom are already integrated out. 
The simplest form of the freeze-out is known as the Cooper-Frye procedure\cite{cooper}.

In this standard procedure of the freeze-out, the fluid elements are considered to flow along smooth streamlines, 
but this assumption is not obvious in viscous fluids.
Viscosity is related to the heat production associated with the motions of constituent particles 
and should comply with the {\it fluctuation-dissipation} theorem.
It is thus natural to consider that the motion of the viscous fluid element is influenced by thermal fluctuation.
Indeed, the Navier-Stokes-Fourier (NSF) equation 
can be derived by the optimization of an action defined for fluctuating fluid elements.
This method is called the stochastic variational method (SVM) \cite{yasue1981}. 
As for other variational approaches to formulate hydrodynamics, see Refs.\ \cite{gonccalves,zambrini} and references therein.
We can show that the velocity fluctuation of the fluid elements is characterized by an uncertainty relation which is analogous to the one in quantum mechanics \cite{koide2018,koide20-2,koide22}.

The velocity fluctuations are different in each hydrodynamic evolution and thus affects the event by event (EbE) analysis of hydrodynamic models 
in relativistic heavy-ion collisions. 
One example is the freeze-out procedure.
The velocity of the fluid element on the freeze-out hypersurface fluctuates and consequently so does the momentum distribution of produced hadrons. 
Thus, we need to introduce the ensemble average for this velocity distribution.
Because of the uncertainty relation mentioned above, 
this velocity fluctuation is associated with the square of the standard deviation of the momentum of the fluid element ($\sigma^{(2)}_p$). 
This is restricted by $\sigma^{(2)}_x$ which reflects the inhomogeneity in the geometrical distribution of the fluid. 
Therefore the standard freeze-out may underestimate the effect of the spatial inhomogeneity of QGP and hence 
the anisotropy of produced hadrons which are measured through the collective flow parameters $v_n$.

In this paper, we discuss the effect of the velocity fluctuation of the fluid element to the collective flow parameter $v_2$.
Unfortunately, the method to estimate the velocity fluctuation is not yet established for relativistic fluids.
Moreover there are several complexities to study quantitative behaviors in relativistic heavy-ion collisions: the setup of initial conditions, 
the choice of the equation of state (EoS) for the QCD matter, the after-burner, for hadronization, etc. 
See for example, Refs.\ \cite{de,hirano} and references therein.
We do not know yet the importance of the velocity fluctuation and hence want to avoid such complexities, which can mask the relation between the velocity fluctuation and anisotropy.
To make this effect on $v_2$ standout,  
we thus choose an initial condition and EoS as simple as possible, and employ the sudden freeze-out scenario with the Cooper-Frye formula in non-relativistic hydrodynamics. 
We show that the presence of such a velocity fluctuation increases the local anisotropy of the momentum distributions of produced hadrons,
and thus the collective flow parameters such as $v_2$ is emphasized.

This paper is organized as follows. In Sec.\ \ref{sec:2}, we briefly summarize the derivation of the NSF equation in SVM. 
The uncertainty relation satisfied for the motion of the fluid element is discussed in Sec.\ \ref{sec:3}. In Sec.\ \ref{sec:4}, 
the modification of the Cooper-Frye formula by the velocity fluctuation is derived in a non-relativistic hydrodynamic model. 
The effect of the velocity fluctuation for $v_2$ is numerically calculated in Sec.\ \ref{sec:5}.
Section \ref{sec:6} is devoted to summary and concluding remarks.

\section{Stochastic dynamics of fluid element} \label{sec:2}

In this section, we briefly summarize the properties of the stochastic motion of the fluid element in the NSF equation, following the review papers \cite{gonccalves,kodama22,koidereview}.

We consider a simple fluid composed of N constituent
particles with mass $m$. The action of the system is given by the sum of the contributions
from each fluid element as
\begin{eqnarray}
I [ {\bf r}(t) ]= \int^{t_f}_{t_i} dt \, \int d^3 {\bf R} \, \rho_0 ({\bf R})  {\cal L} ({\bf r}(t), \dot{\bf r}(t)) 
\, , 
\end{eqnarray}
where $t_i$ and $t_f$ are arbitrary initial and final times and 
\begin{eqnarray}
{\cal L} ({\bf r}(t), \dot{\bf r}(t)) = 
\left[
\frac{m}{2} \left( \frac{d {\bf r}(t)}{dt} \right)^2 - \frac{\varepsilon}{\rho} 
\right]  \, .
\label{eqn:cla-lag}
\end{eqnarray}
where ${\bf r}(t)$ is the Lagrange coordinate of the fluid element and ${\bf R} = {\bf r}(t_i)$ denotes its initial position. 
We introduced the internal energy density $\varepsilon$  
and the number density of the constituent particles $\rho$.
Each fluid element involves a fixed number of constituent particles which is conserved in the time evolution. 
Then the initial distribution of the fluid elements is characterized by $\rho_0({\bf R})$, normalized by $N$,
\begin{eqnarray}
N = \int d^3 {\bf R} \, \rho_0 ({\bf R}) \, .
\end{eqnarray}
Applying the standard variation to this action, we obtain the Euler equation \cite{koide2012}

In SVM, the viscous effect is reproduced from the stochastic motion of the fluid element.
We now consider that 
the dynamics of the fluid is represented by a set of  stochastic Lagrangian coordinates, denoted
by $\widehat{\bf r}(t)$ where $(\widehat{\,\,\,})$ 
denotes a stochastic quantity.
Suppose that the time evolution is determined by the forward stochastic differential equation (SDE),
\begin{eqnarray}
d \widehat{\bf r}(t) = {\bf u}_+ (\widehat{\bf r}(t),t) dt + \sqrt{2\nu } d\widehat{\bf W}_+ (t)  \,\,\, (dt >0)\, . \label{eqn:fsde}
\end{eqnarray}
The standard Wiener process is denoted by $\widehat{\bf W}_+ (t)$ which satisfies 
\begin{eqnarray}
	E\left[d\widehat{\mathbf{W}}_+ (t)\right] = 0 \, , \,\,\,\,
	E\left[d\widehat{W}^i_+ (t)d\widehat{W}^j_+ (t^\prime)\right] = |dt| \delta_{t,t^\prime}\delta_{i,j} \, ,
\label{eqn:wiener}
\end{eqnarray}
where $E[\,\,\,]$ is the ensemble average for the Wiener process.
The intensity of the noise is controlled by the parameter $\nu$ which characterizes thermal fluctuation.
Note that the vector field ${\bf u}_+ ({\bf x},t)$ does not necessarily coincide with the velocity field of the NSF equation.

The NSF equation is obtained by applying the stochastic variation to this action. 
For this, the position of the fluid element in the Lagrangian is replaced by the stochastic quantity.
This replacement is not trivial because the stochastic trajectory is zigzag and then the left and right-hand limits 
of the inclination of stochastic trajectories do not agree even in the vanishing limit of $dt$.
To perform the variational method for stochastic variables, we have to distinguish these two limits.
 We then introduce another stochastic differential equation which describes the backward time evolution.
This is called the backward SDE and defined by 
\begin{eqnarray}
d \widehat{\bf r}(t) = {\bf u}_- (\widehat{\bf r}(t),t) dt + \sqrt{2\nu} d\widehat{\bf W}_- (t)  \,\,\, (dt < 0)\, , \label{eqn:bsde}
\end{eqnarray}
where $\widehat{\bf W}_- (t)$ is another Wiener process which satisfies the same correlation properties as Eq.\ (\ref{eqn:wiener}) 
where $|dt| = - dt$.
Following Nelson \cite{nelson1966}, the left and right-hand limits of the velocity of the fluid element are defined by 
\begin{eqnarray}
D_\pm \widehat{\bf r} (t) &=& \lim_{dt \rightarrow 0\pm} E \left[ \frac{\widehat{\bf r}(t+dt) - {\bf r}(t)}{dt}  \Bigg| {\bf r}(t) \right] = {\bf u}_\pm ({\bf r}(t),t) \, .
\end{eqnarray} 
Here the expectation value is the conditional average for fixing $\widehat{\bf r}(t)$. In the above definitions, we used $\widehat{\bf r}(t)$ is Markovian.

We impose that $\widehat{\bf r}(t)$ satisfies the forward and backward SDE's at the same time. 
From each SDE's we can derive respective Fokker-Planck equations which are required to be equivalent. 
From this, we find the consistency condition, 
\begin{eqnarray}
{\bf u}_+ ({\bf x},t) - {\bf u}_- ({\bf x},t) = 2\nu \nabla \ln \rho ({\bf x},t) \, . \label{eqn:cc}
\end{eqnarray} 
Using this condition, we can show that the two Fokker-Planck equations are reduced to the same equation of continuity for $\rho$, 
\begin{eqnarray}
\partial_t \rho({\bf x},t) = - \nabla \cdot \{\rho({\bf x},t) {\bf v({\bf x},t)} \} \, , \label{eqn:eq_con}
\end{eqnarray}
where
\begin{eqnarray}
{\bf v} ({\bf x},t) = \frac{{\bf u}_+ ({\bf x},t) +{\bf u}_- ({\bf x},t) }{2} \, .\label{eqn:def-vel}
\end{eqnarray}
The velocity in the NSF equation is identified with ${\bf v}$.

The stochastic generalization of Eq.\ (\ref{eqn:cla-lag}) is 
defined by replacing the kinetic term with $D_+$ and $D_-$,
\begin{eqnarray}
{\cal L}_{s t o} \left(\widehat{\mathbf{r}}, \mathrm{D}_+ \widehat{\mathbf{r}}, \mathrm{D}_{-} \widehat{\mathbf{r}}\right)
= \left[ \frac{m}{2}\left(\mathrm{D}_+ \widehat{\mathbf{r}}(t), \mathrm{D}_{-} \widehat{\mathbf{r}}(t)\right) \mathcal{M}
\left(\begin{array}{c}
		\mathrm{D}_+ \widehat{\mathbf{r}}(t) \\
		\mathrm{D}_- \widehat{\mathbf{r}}(t)
	\end{array}\right) 
- \frac{\varepsilon}{\rho} \right] \, ,
\end{eqnarray}
where
\begin{eqnarray}
	\mathcal{M}=\left(\begin{array}{cc}
		\left( \frac{1}{2} + \alpha_{A} \right)\left( \frac{1}{2} + \alpha_{B} \right) & \frac{1}{2}\left( \frac{1}{2} - \alpha_{B} \right) \\
		\frac{1}{2} \left( \frac{1}{2} - \alpha_{B} \right)  & \left( \frac{1}{2} - \alpha_{A} \right)\left( \frac{1}{2} + \alpha_{B} \right)
	\end{array}\right) \, .
\end{eqnarray}
In this replacement, we considered the most general quadratic form of the kinetic term and thus 
two parameters $\alpha_A$ and $\alpha_B$ are introduced. These parameters are absorbed into the definitions of the transport coefficients in hydrodynamics. 
See the discussion in Ref.\ \cite{gonccalves} for details.
Although it is not explicitly expressed, the position of the fluid element in $\varepsilon$ and $\rho$ are given by $\widehat{\bf r}(t)$. 
Note that what we can optimize is only averaged behaviors. 
The stochastic action is thus defined  by 
\begin{eqnarray}
I_{sto} [\widehat{\mathbf{r}}] 
= \int^{t_f}_{t_i} ds\, \int d^3 {\bf R} \, \rho_0 ({\bf R}) 
E[ {\cal L}_{s t o} \left(\widehat{\mathbf{r}}, \mathrm{D}_+ \widehat{\mathbf{r}}, \mathrm{D}_{-} \widehat{\mathbf{r}}\right) ] \, .
\end{eqnarray}

The stochastic variation of this action leads to the following generalized hydrodynamics,
\begin{eqnarray}
(\partial_t + {\bf v}\cdot \nabla) v^{i} = 2\kappa \partial_i \frac{\nabla^2 \sqrt{\rho}}{\sqrt{\rho}} - \frac{1}{m\rho}\partial_i \left\{ P - \left(\mu + \frac{\eta}{D} \right) (\nabla \cdot {\bf v}) \right\}
+ \frac{1}{m\rho} \sum_{j=1}^D \partial_j (\eta E^{ij}) \, ,  \label{eqn:hydro}
\end{eqnarray}
where $P$ is the thermodynamic pressure and 
\begin{eqnarray}
E^{ij} &=& \frac{1}{2} (\partial_i v^j + \partial_j v^{i}) - \frac{1}{D} (\nabla \cdot {\bf v}) \delta_{ij} \, , \\
\eta &=& 2\alpha_A (1+2 \alpha_B) \nu m \rho \, , \\
\kappa &=& 2 \alpha_B \nu^2 \, .
\end{eqnarray}
Here $D$ denotes the number of the spatial dimension.
The second coefficient of viscosity $\mu$ is obtained from the variation of the entropy density \cite{koide2012}. 
The first term on the right-hand side is the additional term which does not appear in the NSF equation. 
This term is considered to be important to discuss the surface effect and quantum fluctuation \cite{koide22,ander,koide19,gonccalves,koide20,kodama22}. 
In the present work, however, we simply ignore this term by setting $\alpha_B = 0$ leading to $\kappa = 0$.
Then we find that the stochastic variation leads to the NSF equation.
It is easy to confirm that the above equation is reduced to the Euler equation in the vanishing limit of $\nu$.

\section{Uncertainty relation in Hydrodynamics} \label{sec:3}

The fluctuation of the fluid element is washed away in the final result of the fluid equation \ (\ref{eqn:hydro}), but 
we can analyze its effect using the solution of the NSF equation and SDE's in SVM.
The vector fields ${\bf u}_\pm ({\bf x},t)$ are obtained from the NSF equation through the relation  
\begin{eqnarray}
{\bf u}_\pm ({\bf x},t) = {\bf  v}({\bf x},t) \pm \nu \nabla \ln \rho ({\bf x},t) \, , \label{eqn:uvrho}
\end{eqnarray}
which is obtained from Eqs.\ (\ref{eqn:cc}) and (\ref{eqn:def-vel}).
This depends on the noise intensity $\nu$ which is an independent quantity of the shear viscosity $\eta$.

The fluctuation mentioned above is intimately related to the hydrodynamic uncertainty relation. 
To see this, we define the standard deviation of the position of the fluid element by  
\begin{eqnarray}
\sigma^{(2)}_{x^{i}} \equiv \lceil (\delta x^{i})^2 \rfloor \, ,  
\end{eqnarray} 
where $\delta f = f({\bf x},t) - \lceil f \rfloor$ and 
\begin{eqnarray}
\lceil f \rfloor = \frac{1}{N} \int d^D {\bf x}\, \rho({\bf x},t) f({\bf x},t) \, .
\end{eqnarray}
For the standard deviation of the momentum of the fluid element, we can introduce two momenta from the stochastic Lagrangian, 
\begin{eqnarray}
{\bf p}_\pm ({\bf x},t)= \left. 2 \frac{\partial L_{sto}}{\partial D_\pm \widehat{\bf r}(t)} \right|_{\widehat{\bf r}(t) = {\bf x}} \, .
\end{eqnarray}
The factor $2$ in the definitions are introduced for a conversion to reproduce the classical result in the vanishing limit of $\nu$ \cite{gonccalves,koide2018,koide22,koide20-2}.
The appearance of the two momenta is attributed to the fact that we cannot define the derivative of the stochastic trajectory uniquely.
Therefore, we assume that the standard deviation of the momentum is defined by the average of the two contributions, 
\begin{eqnarray}
\sigma^{(2)}_{p^{i}} \equiv
\frac{\lceil (\delta p^{i}_+)^2 \rfloor + \lceil (\delta p^{i}_-)^2 \rfloor}{2} \, .
\end{eqnarray}
Using these definitions and the Cauthy-Schwartz inequality, it is straightforward to show 
the uncertainty relation in hydrodynamics
\begin{eqnarray}
\sigma^{(2)}_{x^{i}} \sigma^{(2)}_{p^{j}} 
\ge m^2 \frac{\xi^4}{\nu^2 + \xi^2} \delta_{ij} \, , \label{eqn:ucr}
\end{eqnarray}
where $\xi$ is the kinematic viscosity 
\begin{eqnarray}
\xi = \frac{\eta}{2m\rho} \, .
\end{eqnarray}
This uncertainty relation corresponds to the Kennard inequality in quantum mechanics and the shear viscosity plays a role of the Planck constant.
The more general expressions of the uncertainty relations in hydrodynamics are shown in Refs.\ \cite{koide2018,gonccalves,koide22}.
This inequality indicates that $\sigma^{(2)}_{p^{i}}$ is affected by the spatial inhomogeneity of the fluid.

\section{Modification of Freeze-out procedure} \label{sec:4}

Let us consider the non-relativistic approximation of the standard freeze-out procedure.
The fluid picture loose its validity when its energy density becomes lower than a critical value 
and afterwards the fluid is replaced with the streaming of non-interacting produced hadrons.
This critical energy density is normally characterized by the freeze-out temperature $T_{FO}$.
When the temperature of the fluid element is cooled down to $T_{FO}$, we replace the element with an ensemble of the ideal gas of observed hadrons. 
Then the momentum distribution of the hadrons is calculated by using the Cooper-Frye formula, 
\begin{eqnarray}
\frac{E}{c} \frac{d^{D} N_{i}}{d \mathbf{p}^{D}}=\int_{\Sigma} d \sigma \, n_{\mu} p^{\mu} f(x, p) \, ,\label{cf1} 
\end{eqnarray}
where $c$ is the speed of light, $f(x,p)$ is the phase space distribution of produced hadrons and 
the four momentum $p^{\mu} = (\sqrt{\um^2 c^2 + {\bf p}^2 }, {\bf p})$ with $\um$ being the mass of a hadron.
The integral is implemented on the $D+1$ dimensional hypersurface which is defined by the space-time positions of the fluid elements at $T_{FO}$.
This hypersurface is called the freeze-out hypersurface.
The normal vector of the $D+1$ dimensional surface element $d\sigma$ is denoted by $n_\mu$.

Note that the normal vector of the freeze-out hypersurface is parallel to the gradient of the temperature. 
The time component of the normal vector is proportional to $c^{-1}$ and thus the normal vector behaves as the space-like vector in the non-relativistic limit,  
\begin{eqnarray}
n_\mu = \frac{1}{|\nabla T|} (0, -\nabla T) \, .
\end{eqnarray}
In the calculation of the product $p^\mu n_\mu$, however, we cannot drop the time component because $p^{0} \approx \um c$, and thus 
we find 
\begin{eqnarray}
p^\mu n_\mu \approx  \frac{\um \partial_{t} T-\mathbf{p} \cdot \nabla T}{|\nabla T|} \, .
\end{eqnarray}
Using these results, the non-relativistic limit of the Cooper-Frye formula is expressed as 
\begin{eqnarray}
\frac{d^{D} N}{d p^{D}}= \int_{\Sigma} d \sigma \left(\frac{\partial_{t} T}{|\nabla T|}-\frac{\mathbf{p} \cdot \mathbf{n}}{\um}\right) f_{boltz}({\bf x},{\bf  p}; {\bf v}_f) \, ,\label{cf}
\end{eqnarray}
where $f_{boltz}({\bf x},{\bf  p}; {\bf v}_f)$ is the Boltzmann distribution, 
\begin{eqnarray}
 f_{boltz}({\bf x},{\bf  p}; {\bf v}_f)= \frac{g}{(2\pi\hbar)^D} e^{- \frac{({\bf p} - \um {\bf v}_f )^2}{k_B (2\um)T_{FO}} } \, ,
\end{eqnarray}
with $k_B$ and $g$ being the Boltzmann constant and the degeneracy factor, respectively.
Here ${\bf v}_f$ is the velocity of the fluid element on the freeze-out hypersurface. 
When the fluid element is deterministic and flows along the streamline, ${\bf v}_f $ is given by the fluid velocity  ${\bf v}({\bf x},t)$. 
This is the standard freeze-out procedure.
We however know that the trajectory of the fluid element fluctuates satisfying the uncertainty relation. 
Thus to take this effect into account, we estimate the fluctuation of ${\bf v}_f $.

This velocity fluctuation is not equivalent to the momentum fluctuation characterized by $\sigma^{(2)}_{p^{i}}$.
Because the viscosity is the velocity-dependent force and thus the momenta defined above are not parallel to the velocity.
This is similar to the relation between the momentum of the electric charge and the Lorentz force.  
The fluctuation of the velocity of the fluid element at $\widehat{\bf r}(t) ={\bf x}$ 
is defined by
\begin{eqnarray}
I^{i} ({\bf x},t) 
&=& \frac{1}{2} \left\{
\lim_{dt \rightarrow 0_+} E\left[ \left(\frac{d\widehat{r}^{i}(t)}{dt} 
- v^{i}(\widehat{\bf r}(t) ,t  \right)^2  \Bigg| \widehat{\bf r}(t) = {\bf x}\right] 
\right. \nonumber \\
&& \left.  +
\lim_{dt \rightarrow 0_-} E\left[ \left(\frac{d\widehat{r}^{i}(t)}{dt} 
- v^{i}(\widehat{\bf r}(t) ,t )  \right)^2  \Bigg| \widehat{\bf r}(t) = {\bf x}\right]
\right\} \, .
\end{eqnarray}
Again, we used the conditional average.
Substituting the forward and backward SDE's and using the consistency condition (\ref{eqn:cc}), 
this correlation function is given by
\begin{eqnarray}
I^{i} ({\bf x},t) = (\nu \partial_i \ln \rho ({\bf x},t))^2 \, .\label{eqn:std}
\end{eqnarray}

Note that $\left[ d\widehat{r}^{i}(t)/dt - u^{i}_\pm (\widehat{\bf r}(t) ,t ) \right]$ are proportional to the inclination of the Wiener process, as is seen from Eqs.\ (\ref{eqn:fsde}) and (\ref{eqn:bsde}), and thus the distribution of these quantities are characterized by Gaussian functions. 
This is not applicable to $\left[d\widehat{r}^{i}(t)/dt - v^{i}_\pm (\widehat{\bf r}(t) ,t ) \right]$ 
because ${\bf u}_\pm ({\bf x},t) \neq {\bf v} ({\bf x},t)$ in general. 
We, however, assume that the deviation from the Gaussian form is small and then the probability distribution is approximately given by 
\begin{eqnarray}
n({\bf x}, {\bf v}_f, t) = \prod^D_{i=1} \sqrt{\frac{1}{2I^{i}({\bf x},t)}} e^{-(v^{i}_f - v^{i}({\bf x},t))^2/(2I^{i}({\bf x},t))} \, ,
\end{eqnarray}
where we can confirm 
\begin{eqnarray}
\int d^D {\bf v}_f \, n({\bf x}, {\bf v}_f, t) v^{i}_f &=& v^{i} ({\bf x},t) \, ,\\
\int d^D {\bf v}_f \, n({\bf x}, {\bf v}_f, t) (v^{i}_f - v^{i} ({\bf x},t) )^2 &=& I^{i} ({\bf x},t)  \, .
\end{eqnarray}

Due to the fluctuation of ${\bf v}_f$,  
the Cooper-Frye formula is averaged with this distribution of the velocity fluctuation.
In the end, the modified Cooper-Frye formula is given by 
\begin{eqnarray}
\frac{d^{D} N}{d p^{D}}
&=& 
\int d^D {\bf v}_f \, n({\bf x}, {\bf v}_f, t) \int_{\Sigma} d \sigma \left(\frac{\partial_{t} T}{|\nabla T|}-\frac{\mathbf{p} \cdot \mathbf{n}}{\um}\right) f_{boltz}({\bf x},{\bf  p}; {\bf v}_f) \nonumber \\
&=&
 \int_{\Sigma} d \sigma \left(\frac{\partial_{t} T}{|\nabla T|}-\frac{\mathbf{p} \cdot \mathbf{n}}{\um}\right) f_{eff}({\bf x},{\bf  p}) \, ,
\label{eqn:mcf_formula}
\end{eqnarray}
where the effective Boltzmann distribution $ f_{eff}({\bf x},{\bf  p}) $ is introduced.
This distribution is characterized by ``anisotropic temperature index'' denoted by $(T^{1}_{AI}, \cdots, T^{D}_{AI})$,
\begin{eqnarray}
f_{eff}({\bf x},{\bf  p}) &=& \frac{g}{(2\pi\hbar)^D} \prod_{i=1}^{D}\sqrt{\frac{\um}{2k_B T^{i}_{AI} ({\bf x},t)}}
 e^{-\frac{\um}{2k_B T^{i}_{AI} ({\bf x},t)}\left(\frac{p^i}{\um}-v^i ({\bf x},t) \right)^2} \, , \\
T^{i}_{AI} ({\bf x},t) &=& T_{FO} + \frac{\um}{2k_B} I^{i} ({\bf x},t) \,\,\,\,\     (i=1,\cdots, D)\, .\label{temeff}
\end{eqnarray}
One can see that the anisotropic temperature index behaves as the temperature for each spatial component.

\section{Illustrative example} \label{sec:5}

In the following, we employ the natural unit where $\hbar =1$, $c= 1$ and $k_B =1$.
Using the modified Cooper-Frye formula (\ref{eqn:mcf_formula}), we study the effect of the velocity fluctuation to the collective flow parameter $v_2$ of the produced hadrons with a mass of the same order of magnitude as that of pions, $\um_\pi = 0.7$ ${\rm fm}^{-1}$.
We consider the non-relativistic fluid described by the NSF equation. For simplicity, we ignore the bulk viscosity by setting $\mu = - \frac{\eta}{D}$ and 
consider the 2+1 dimensional system, $D=2$.
Therefore we solve the coupled equations, 
\begin{eqnarray}
(\partial_t + {\bf v}\cdot \nabla) v^{i} &=& - \frac{1}{\um_\pi \rho}\partial_i P   + \frac{1}{\um_\pi \rho} \sum_{j=1}^2 \partial_j (\eta E^{ij}) \, ,  \\
\left( \partial_t + {\bf v} \cdot \nabla \right) \varepsilon &=& -\sum_{i,j =1}^{2} (P\delta^{ij}-\eta E^{ij})\partial_j v^i \, , \label{eqn:eq_ener}
\end{eqnarray}
together with the equation of continuity (\ref{eqn:eq_con}). 
For the shear viscosity, we apply 
\begin{eqnarray}
\frac{\eta}{2\rho} = \frac{\hbar}{8\pi} \, .
\end{eqnarray}
As is discussed in Ref.\ \cite{koide22}, this can be regarded as the Kovtun-Son-Starients (KSS) bound \cite{son} of the shear viscosity for the NSF equation. 
The corresponding Reynolds number is given by ${\rm Re} \approx 8$ in choosing 
the typical scales of length and velocity by $\sim 1$ fm and $c$, respectively.
The initial 2-dimensional particle density and the initial energy density are, respectively, chosen as 
\begin{eqnarray}
\rho ({\bf x},0) &=& \rho_c \exp\left( -\frac{x^2}{R^2_x} - \frac{y^2}{R^2_y} \right) \, {\rm fm}^{-2}\, ,\\
\varepsilon({\bf x}, 0) &=& a  ( \rho(\mathbf{x},0))^{5/3} \, {\rm fm}^{-3}\, ,
\end{eqnarray}
with $(R_x, R_y ) = ( 4/3 , 3 ) \, {\rm fm}$. 
We take EoS given by  
\begin{eqnarray}
P = b \, \rho^{5/3} \, {\rm fm^{-3}}\, ,
\end{eqnarray}
and then the temperature is 
\begin{eqnarray}
T  =b \, \rho^{2/3} \, {\rm fm}^{-1} \, .
\end{eqnarray}
These constants in the present calculation are chosen by  
\begin{eqnarray}
\rho_c &=& 7.9 \, {\rm fm}^{-2} \, ,\\
a &=& 0.167 \, {\rm fm}^{-1/3} \, ,\\
b &=& 0.11 \, {\rm fm}^{-1/3} \, .
\end{eqnarray}
In this choice,  the maximum initial temperature is $\sim 0.5 \, {\rm fm}^{-1}$.
The freeze-out temperature is set as
\begin{eqnarray}
T_{FO} = 0.4 \, {\rm fm}^{-1} \, .
\end{eqnarray}

\begin{figure}[H]
\begin{tabular}{cc}
\begin{minipage}[t]{0.4\hsize}
\includegraphics[scale=0.3]{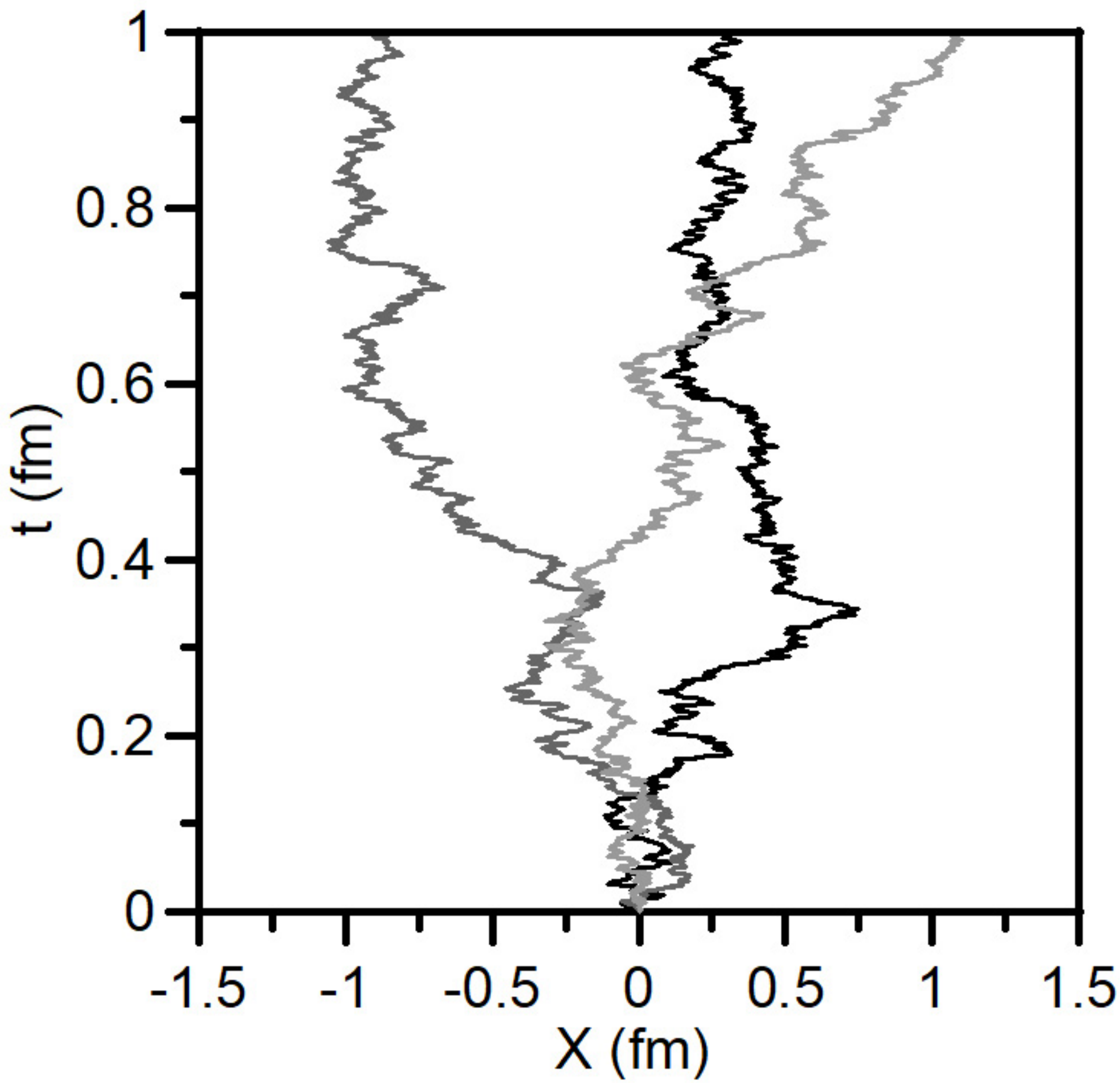}
\end{minipage} &
\hspace{1cm}
\begin{minipage}[t]{0.4\hsize}
\includegraphics[scale=0.3]{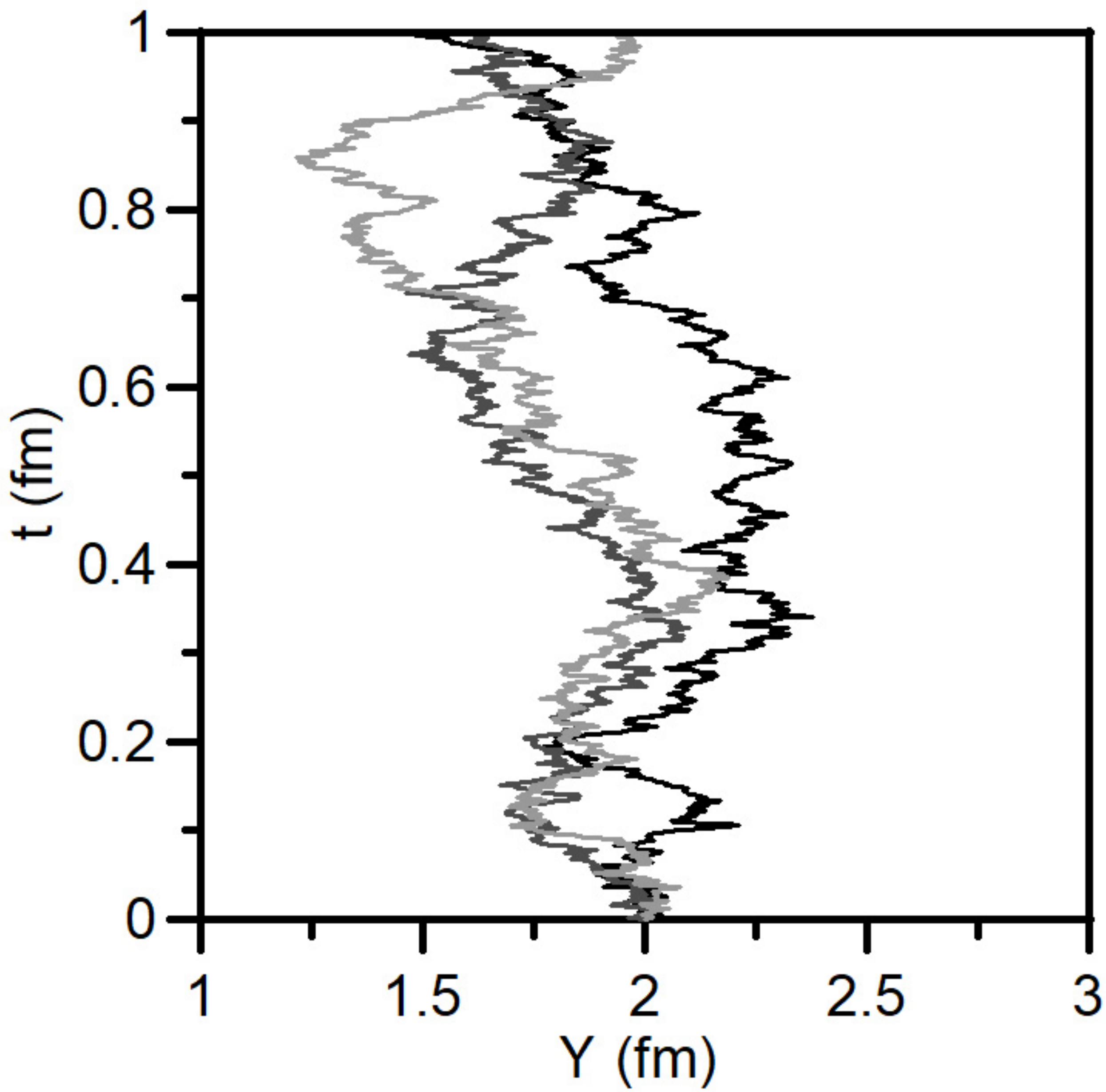}
\end{minipage} \\
\begin{minipage}[t]{0.4\hsize}
\includegraphics[scale=0.3]{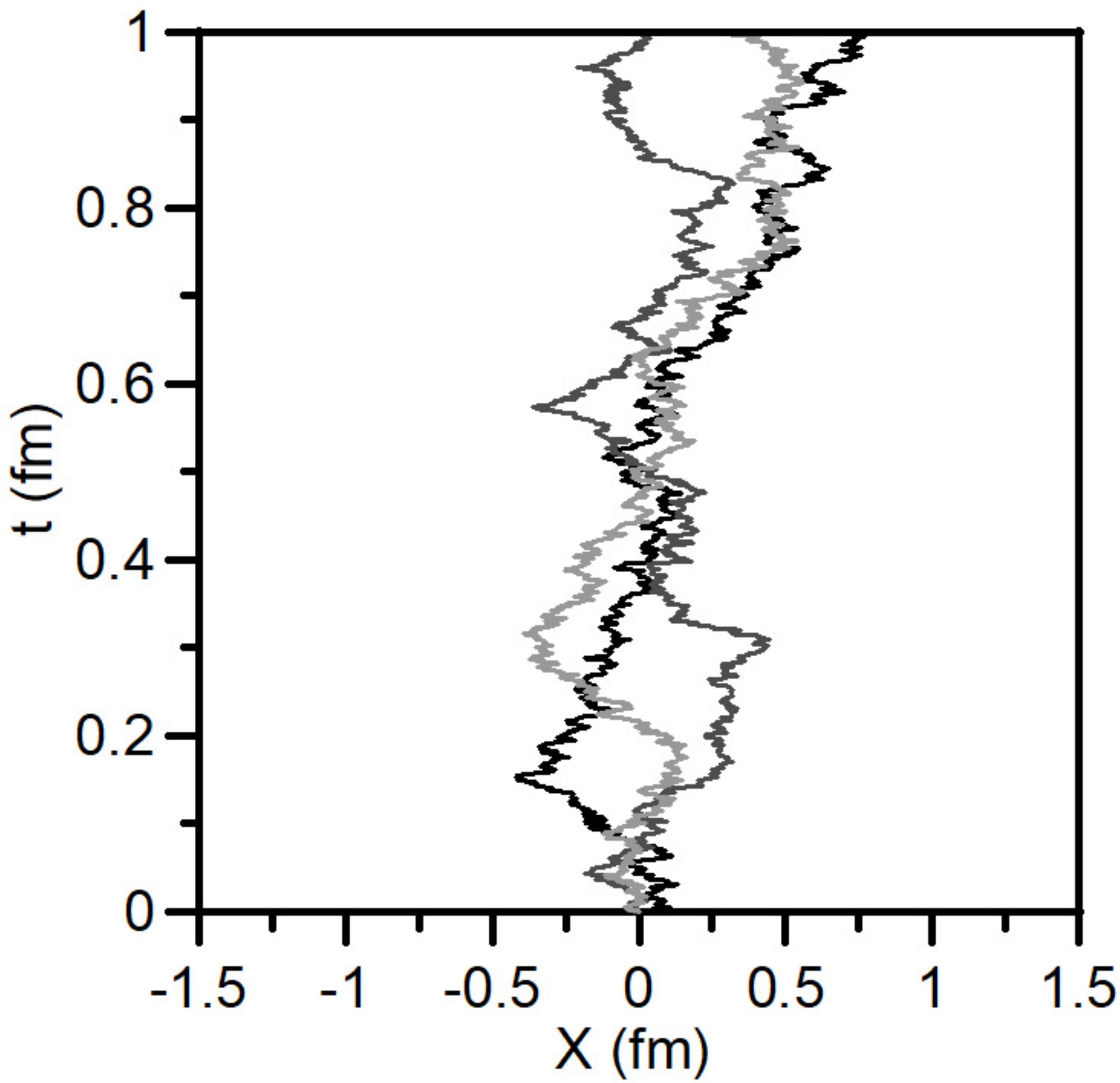}
\end{minipage} &
\hspace{1cm}
\begin{minipage}[t]{0.4\hsize}
\includegraphics[scale=0.3]{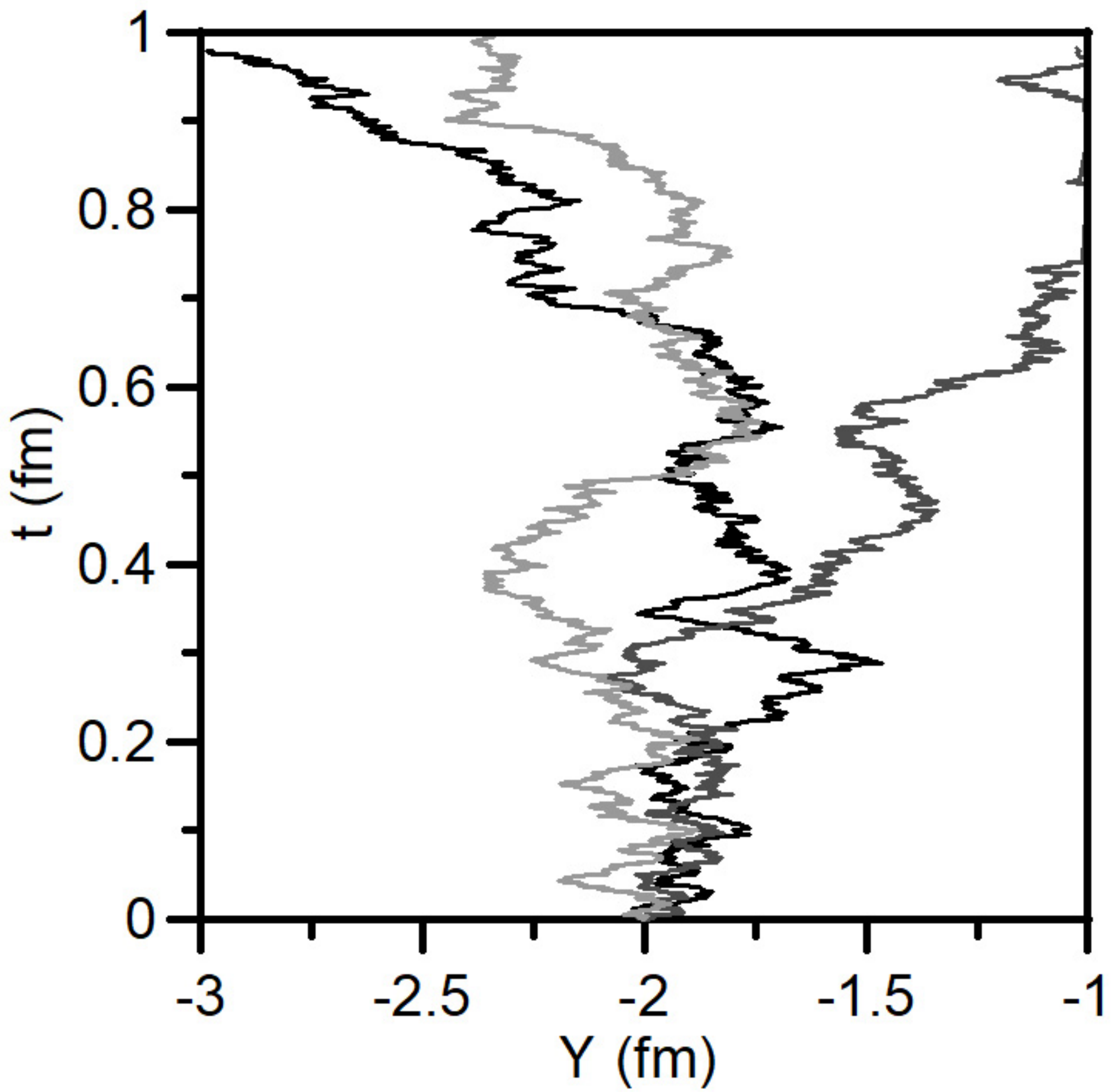}
\end{minipage} \\
\end{tabular}
\caption{
The upper figures show the trajectories of the fluid element for three different stochastic events starting from the same initial position $(x(0),y(0)) = (0,2)\, {\rm fm}$, 
which are obtained  by solving the forward SDE (\ref{eqn:fsde}). The left and right panels correspond to the $x$ and $y$ components of the trajectories, respectively. 
The lower figures gives the trajectories from the initial position $(x(0),y(0)) = (0,-2)\, {\rm fm}$.
The left and right panels correspond to the $x$ and $y$ components of the trajectories, respectively. 
For these figures, we used $\nu = 0.2 \, {\rm fm} \, $.
}
\label{fig:1}
\end{figure}

It is worth observing the examples of the stochastic motion of the viscous fluid element, which are shown in Fig.\  \ref{fig:1}. 
The upper figures show the trajectories of the fluid element for three different stochastic events starting from the same initial position $(x(0),y(0)) = (0,2)\, {\rm fm}$. 
The left and right panels correspond to the $x$ and $y$ components of the trajectories, respectively. 
The lower figures gives the trajectories from the initial position $(x(0),y(0)) = (0,-2)\, {\rm fm}$.
These trajectories are obtained by solving the forward SDE (\ref{eqn:fsde}) applying the Euler-Maruyama method with $dt=0.001\, {\rm fm}$.
The vector field $\mathbf{u}_+$ is given by Eq.\ (\ref{eqn:uvrho}).
The intensity is chosen by $\nu=0.2 \, {\rm fm}$.
One can clearly see the fluctuating motions of the fluid elements.
The distribution of the trajectories of all fluid elements should reproduces $\rho ({\bf x},t)$, 
although the numerical check is not an easy task due to the necessity of an enormous number of events.

\begin{figure}[htp!]
	\centering
	\includegraphics[scale=0.25]{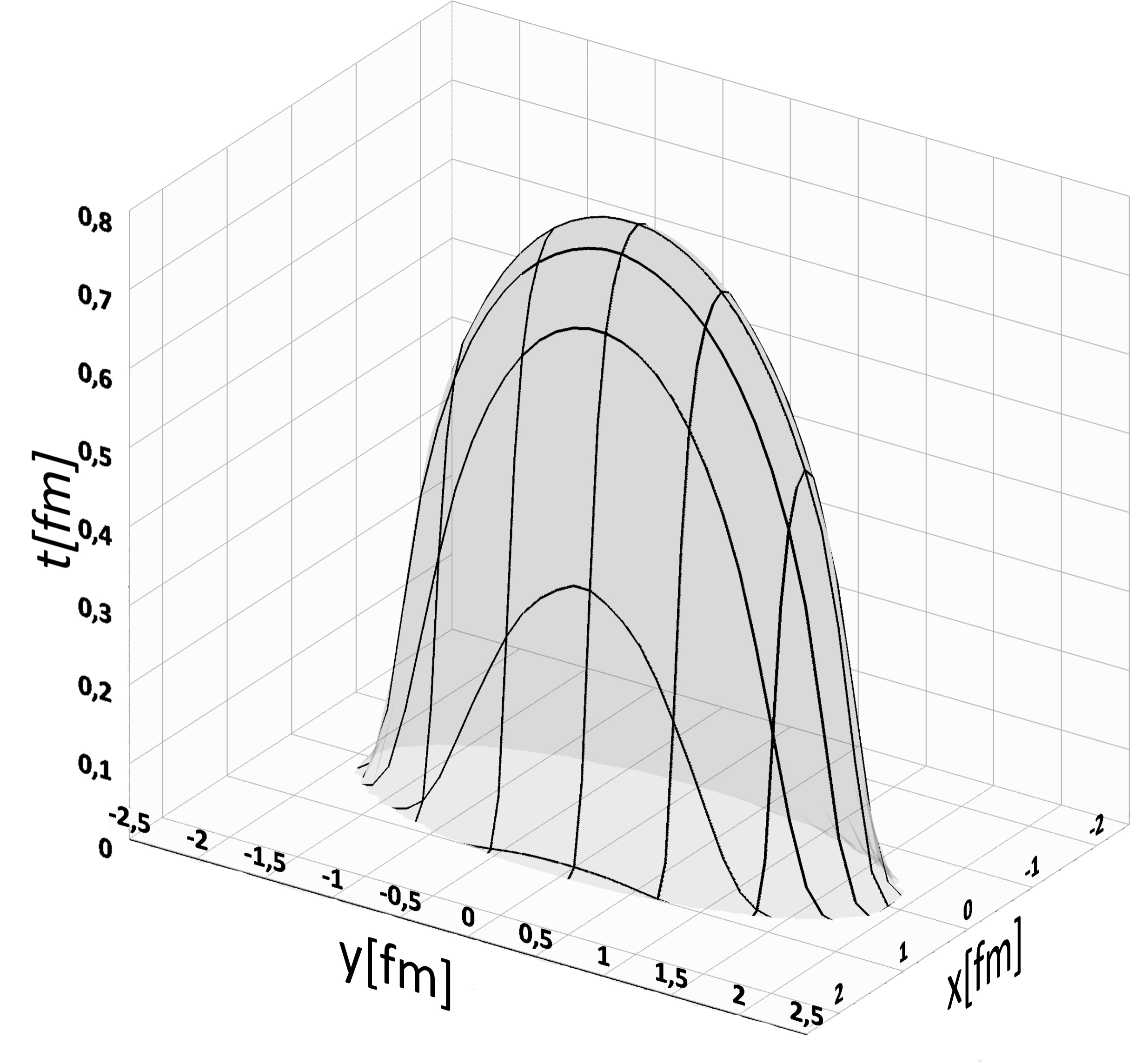}
	\caption{Freeze-out hypersurface in the 2+1 NSF equation}
	\label{fig:freezesurface}
\end{figure}

The freeze-out hypersurface in our simulation is shown in Fig.\ \ref{fig:freezesurface}.
The form is analogous to the shape of a fingertip.
In the present simulation, all fluid elements pass through the hypersurface in the short time period $< 1 \, fm$, but 
the fluid is accelerated around $30 \, \%$ of the speed of light on the freeze-out hypersurface.

\begin{figure}[th!]
	\centering
	\includegraphics[scale=0.3]{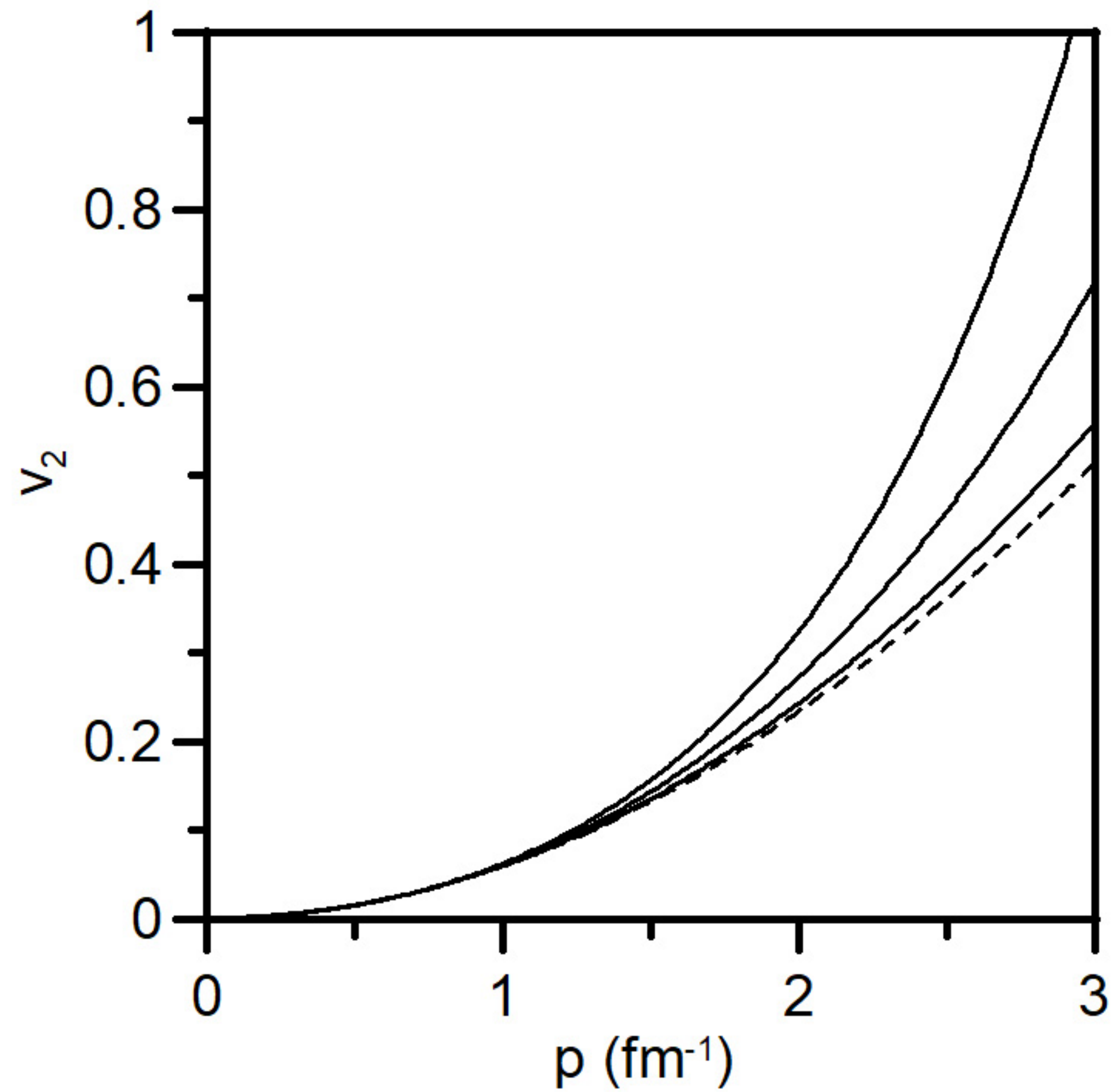}
	\caption{The momentum dependences of $v_2$ are plotted by the solid lines, which correspond to $\nu = 0.2 \, {\rm fm}$, $0.4 \, {\rm fm}$ and $0.6 \,{\rm fm}$ in order from bottom to top.
For the sake of comparison, $v_2$ for the standard Cooper-Frye formula (that is, $\nu=0$) is denoted by the dotted line.  }
\label{fig:v2}
\end{figure}

The collective flow parameter $v_2$ is defined by 
\begin{eqnarray}
	v_{2}(p)=\frac{\int_{0}^{2 \pi} d \phi\, \cos (2\phi) \frac{d^{2} N}{d p^{2}}}{\int_{0}^{2 \pi} d \phi\, \frac{d^{2} N}{d p^{2}}} \, ,\label{v2}
\end{eqnarray} 
where $(p_x,p_y) = (p\cos \phi, p\sin \phi)$ and the hadron spectrum $d^{2} N/d p^{2}$ is given by Eq.\ (\ref{eqn:mcf_formula}).
This modified spectrum depends on another parameter $\nu$ which can be chosen independently of the parameters in hydrodynamics. 
It should be noted that Eq.\ (\ref{eqn:mcf_formula}) is reduced to the standard Cooper-Frye formula in the vanishing limit of $\nu$.
As is seen from the forward and backward SDE's, $\nu$ is associated with the diffusion process in the fluid. 
In fact, $\nu$ coincides with the diffusion coefficient in the Fokker-Planck equation. 
Therefore, in the application to relativistic heavy-ion collisions, this quantity will be given by the diffusion coefficient in QCD.
In this paper, however, we use the non-relativistic toy model and are interested only in the qualitative effect of the velocity fluctuation. 
For this purpose, we consider three values of $\nu$ to see the $\nu$ dependence in $v_2$. 
In Fig.\ \ref{fig:v2}, the momentum dependences of $v_2$ are plotted by the solid lines, which correspond to $\nu = 0.2$  fm, $0.4$ fm and $0.6$ fm in order from bottom to top.
For the sake of comparison, $v_2$ for the standard Cooper-Frye formula (that is, $\nu=0$) is denoted by the dotted line. 
All lines are increasing function of the momentum.
We further observe that the increase of $v_2$ is enhanced for larger $\nu$. 
This is an interesting behavior because one would normally expect that fluctuation has a tendency to homogenize systems.

\begin{figure}[h]
\begin{tabular}{cc}
\begin{minipage}[t]{0.45\hsize}
\includegraphics[scale=0.3]{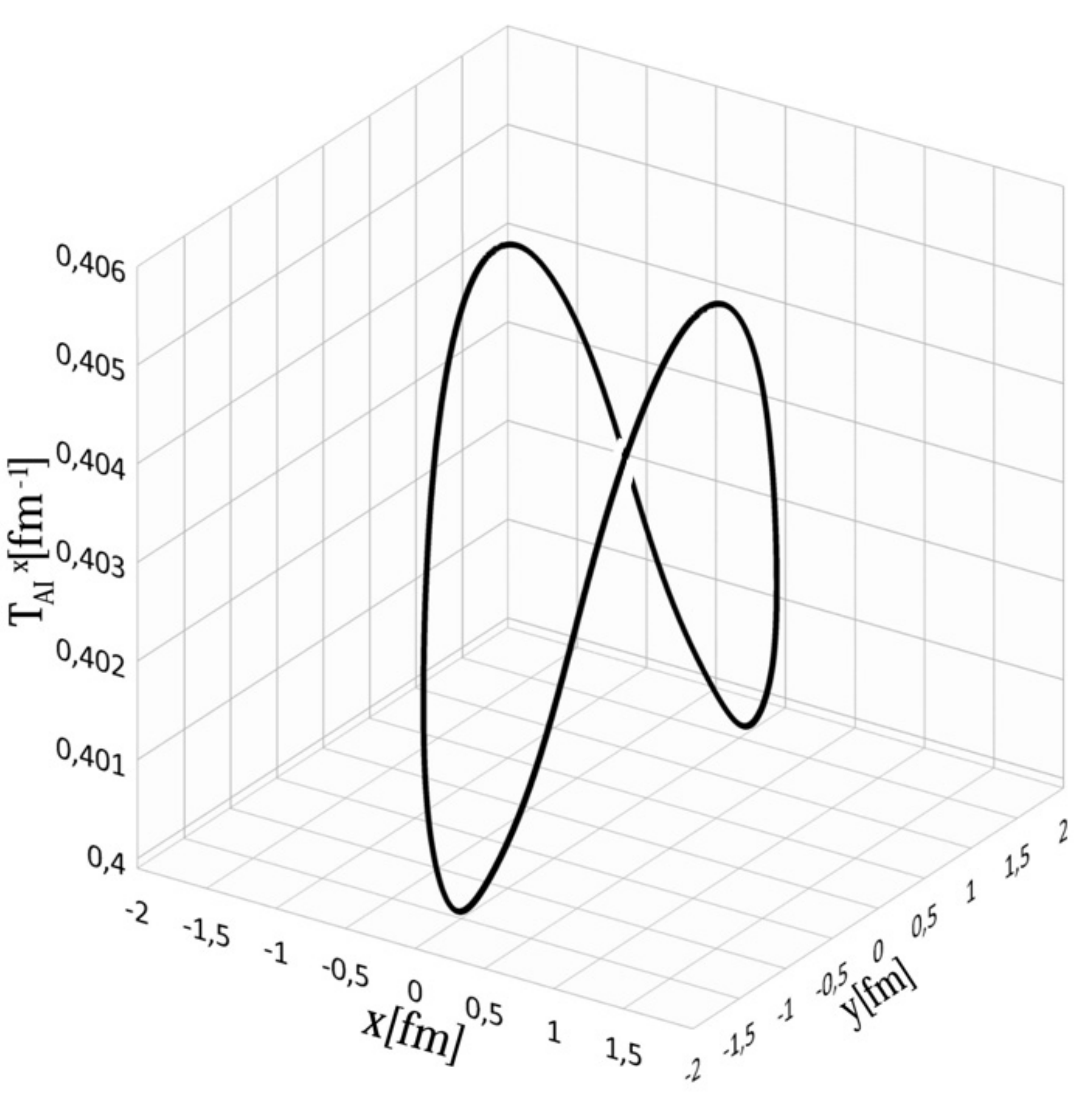}
\end{minipage} &
\begin{minipage}[t]{0.45\hsize}
\includegraphics[scale=0.3]{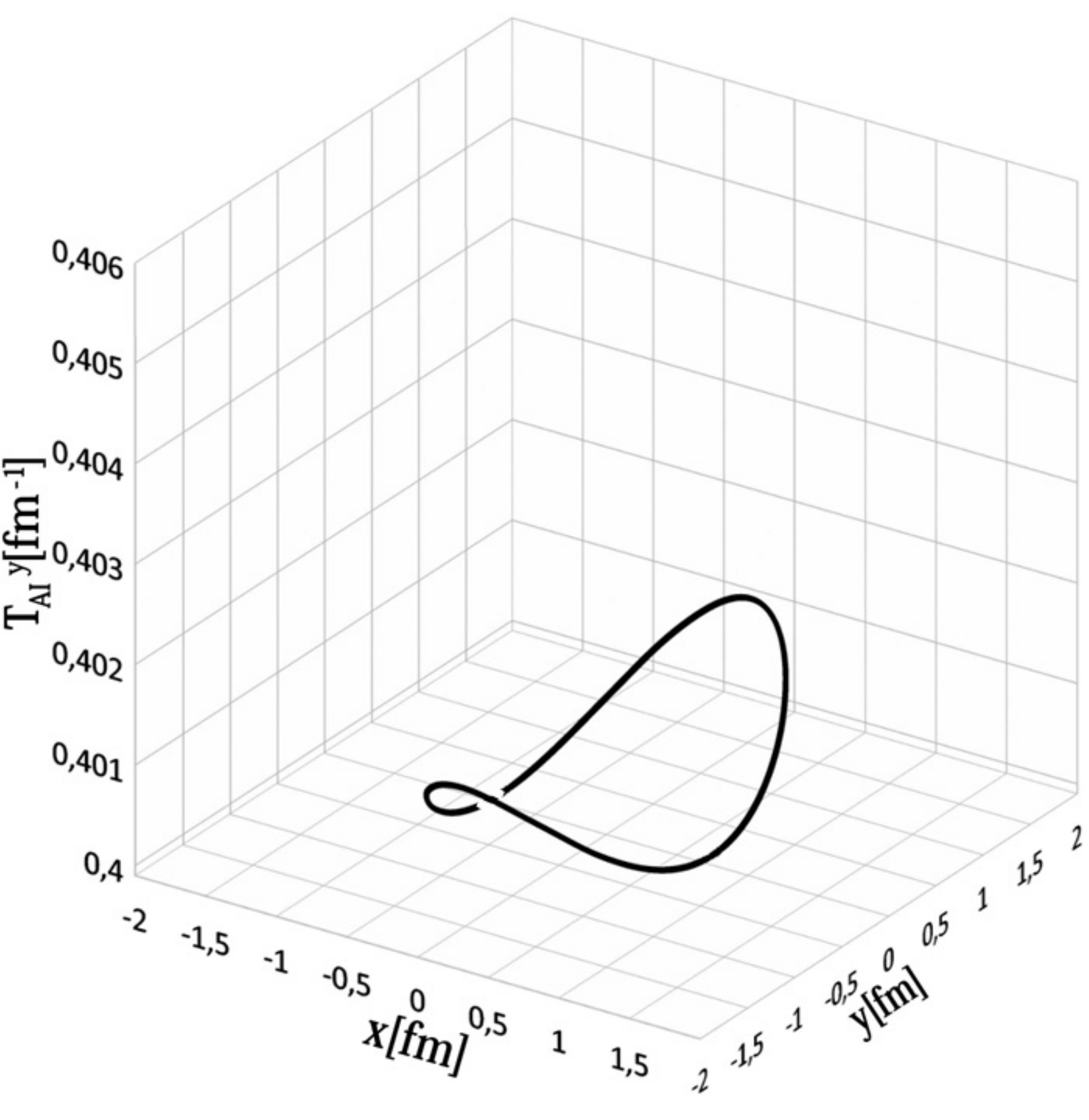}
%\subcaption{The discrepancy in Fig.\ \ref{fig:variableQ2}.}
%\label{fig:disQ2}
\end{minipage} 
\end{tabular}
\caption{
The  anisotropic temperature indices are shown for $\nu = 0.2 \, {\rm fm}$ at a fixed time $t=0.4\, {\rm fm}$. The left and right panels are the $x$ and $y$ components of the temperature, respectively. The $x$ ($y$) component $T^{x}_{AI}$ ($T^{y}_{AI}$) becomes maximum for larger $|x|$ ($|y|$).}
	\label{fig:effectivet}
\end{figure}

This intriguing result is attributed to the behavior of the anisotropic temperature index (\ref{temeff}).
In Fig.\ \ref{fig:effectivet}, the  anisotropic temperature indices on the freeze-out hypersurface 
are shown for $\nu = 0.2$ fm at a fixed time $t=0.4$ fm.
The left and right panels are the $x$ and $y$ components of the anisotropic temperature index, respectively.
The $x$ component $T^{x}_{AI}$ becomes maximum for larger $|x|$.
This is because $(\partial_x \ln \rho)^2$ is maximized for the largest values of $|x|$ in the present simulation. 
The variation of the anisotropic temperature index is the order of a few percent at maximum.
Nevertheless, this small change enhances the production of hadrons with larger value of $p^{x}$, sensitively. 
The same discussion is applied to the $y$ component $T^{y}_{AI}$.
The change of $T^{x}_{AI}$ is larger than that of $T^{y}_{AI}$, and hence the ratio $p^x/p^y$ of the produced hadrons increases for larger $\nu$.
This is the reason why $v_2$ increases by the influence of the velocity fluctuation.

\section{Concluding Remarks} \label{sec:6}

In this paper, we studied the effect of the velocity fluctuation in the behavior of the collective flow parameter $v_2$.
In the standard hadronization scenario of relativistic hydrodynamic models, the fluid elements are assumed to flow along smooth streamlines.
In viscous fluids, however, the motion of the fluid element fluctuates thermally and this fluctuation is characterized by 
an uncertainty relation which is analogous to the one in quantum mechanics. 
This velocity fluctuation reflects the local spatial anisotropy of the fluid distribution 
and thus the final hadron spectrum is modified.

In the present non-relativistic toy model, the velocity fluctuation along a certain spatial direction is 
characterized by the corresponding gradient of the fluid distribution.
Therefore the velocity fluctuation becomes significant in the domain where the distribution of the fluid changes rapidly.
This effect is characterized by 
the anisotropic temperature index, which is a parameter to represent 
anisotropy in the momentum distribution of the produced hadrons. 
To be more specific, due to the velocity fluctuation, 
the temperature of each spatial component in the Cooper-Frye formula is replaced with the corresponding component of 
the anisotropy temperature index.
As a result, we found that the collective flow parameter $v_2$ increases in the present initial condition.
Our result is not applicable to see the quantitative influence to 
the physics of relativistic heavy-ion collisions, but 
the observed behavior deserves attention because we normally expect that fluctuation homogenizes systems 
and thus does not contribute to the enhancement of anisotropy such as $v_2$.

We used the simple Gaussian distribution for the initial condition.
When we apply the present analysis to the more complex geometrical distribution, 
we will gain a new insight from correlation of the collective flow parameters.
For example, the distinction of finite $v_{3}$ flows between a global triangular anisotropy and the presence of hot
spot in the peripheral region is not clear \cite{PeriferalTube1,PeriferalTube2}. 
The velocity fluctuation may clarify the difference of the $p_T$ dependence in $v_3$ and the correlations among $v_n$ in these scenarios.

The velocity fluctuation is attributed to the non-differentiability of the trajectory of the fluid element.
A similar non-differentiability will appear in relativistic fluids of each collisional event.
In fact, the Dirac equation can be interpreted as the stochastic particle motion satisfying the Poisson process \cite{feynman,kac,gaveau}. 
However, a particle-like localized state is known to be difficult to define in relativistic quantum systems \cite{newton-wigner}.
Indeed, the elementary degrees of freedom are described by fields, not by particles in relativistic quantum field theory.
Therefore, to apply the present idea to relativistic fluids, 
it might be necessary to develop a new hydrodynamic model based on a field theory \cite{kknew}. 
In the present framework, the fluid element is stochastic but hydrodynamics itself is deterministic. 
It is thus interesting to consider how the velocity fluctuation is modified when hydrodynamics becomes stochastic \cite{kapsta}.  
These generalizations are left as future tasks.

\vspace*{1cm}
The authors acknowledge the financial supports by CNPq (No.\ 305654/2021-7, \, 303246/2019-7),
FAPERJ (E-26/010.002107/2019, E-26/010/000077/2018) and CAPES. 
A part of this work has been done under the project INCT-Nuclear Physics and Applications (No.\ 464898/2014-5).

\end{document}